\documentclass[12pt]{article}

\usepackage{graphicx}

\title{A Common Origin of Neutralino Stars and Supermassive Black Holes}

\author{V. I. Dokuchaev\thanks{dokuchaev@inr.npd.ac.ru},
        ~Yu. N. Eroshenko\thanks{erosh@ns.ufn.ru} \\
{\small\sl
Institute for Nuclear Research of the Russian Academy of Sciences, Moscow}}

\begin{document}
\date{}
\maketitle

\begin{abstract}
To account for the microlensing events observed in the Galactic
halo, Gurevich, Zybin, and Sirota have proposed a model of gravitationally
bound, noncompact objects with masses of $\sim(0.01\div1)M_{\odot}$.
These objects are
formed in the expanding Universe from adiabatic density perturbations and
consist of weakly interacting particles of dark matter, for example,
neutralinos. They assumed the perturbation spectrum on some small scale to
have a distinct peak. We show that the existence of this peak would
inevitably give rise to a large number of primordial black holes (PBHs) with
masses of $\sim10^5M_{\odot}$ at the radiation-dominated evolutionary
stage of the Universe. Constraints on the coefficient of nonlinear
contraction and on the compactness parameter of noncompact objects were
derived from constraints on the PBH number density. We show that noncompact
objects can serve as gravitational lenses only at a large PBH formation
threshold, $\delta_{\mathrm{c}}>0.5$,
or if noncompact objects are formed from entropic
density perturbations.
\end{abstract}

\section{INTRODUCTION}

Dark objects with masses of $\sim(0.01\div1)M_{\odot}$ were detected in the
Galactic halo when the microlensing of stars from the Large Magellanic Cloud
was observed. Brown and white dwarfs and Jupiter-like planets were proposed
as possible baryonic candidates. According to the latest data [1], dark
objects must account for about 20\%
of the hidden mass in the Galactic halo.
However, the theory of primordial nucleosynthesis imposes much more
stringent constraints on the number of such baryonic objects.
Gravitationally bound, noncompact objects, which are also called neutralino
stars, were considered among nonbaryonic candidates; these objects can
explain some microlensing events with evidence of gravitational lenses being
noncompact [2, 3]. The hypothetical noncompact objects are the lightest
objects in the hierarchical large-scale structure of the Universe. In the
model of Gurevich, Zybin, and Sirota [3], these are formed immediately after
the passage of the Universe to the dust stage.

To reconcile the parameters of noncompact objects with data for the observed
microlensing events, Gurevich et al. [3] assumed the cosmological density
perturbation spectrum on some small scale to increase sharply with an rms
fluctuation of the order of 1 at the peak. For adiabatic perturbations, rms
fluctuations of the order of 0.05 correspond to these fluctuations at the
radiation-dominated stage on the cosmological horizon scale. As was shown
in [4, 5], fluctuations with such an rms value give rise to primordial black
holes (PBHs) at the radiation-dominated evolutionary stage of the Universe.
As we will see below, the mass of the forming PBHs exceeds the mass of
noncompact objects by several orders of magnitude. The large difference
between the masses of noncompact objects and PBHs stems from the fact that
the mass of the radiation contained in a fixed comoving volume at the
radiation-dominated stage is much larger than the mass of cold dark matter
(CDM) in the same volume at the dust stage.

If the power spectrum of primordial cosmological perturbations is a power law
with an index $n > 1$, then PBHs are formed in a wide range of masses. If,
however, the spectrum has a peak on some scale, then PBHs are formed mostly
in a narrow range of masses, near the mass that corresponds to this peak. A
sharp peak emerges in the fluctuation spectrum if the inflationary potential
$V(\phi)$ has a flat segment [6, 7].
Indeed, if the derivative at some value of
the scalar field is
$V'=dV(\phi)/d\phi\to0$,
then a peak emerges in the perturbation spectrum on the corresponding scale,
because the mean density perturbation on the horizon scale is
$\delta_{\mathrm{H}}\sim M_{\mathrm{Pl}}^{-3}V^{3/2}/V'$,
where $M_{Pl}$ is the Planck mass. A similar effect is achieved in inflationary
models with several scalar fields [8, 9]. In this case, the spectrum outside
the peak can have an ordinary shape, for example, it can be a
Harrison--Zel'dovich spectrum, and can give rise to galaxies, their clusters
and  superclusters according to standard scenarios.

Certain evidence for a deviation of the initial perturbation spectrum
from a simple power-law shape, more specifically, for a spectral break
near large scales, $k\sim(0.06-0.6)h$~Mpc$^{-1}$, was obtained in the
counting of APM galaxies and in the Boomerang and Maxima cosmic microwave
background (CMB) anisotropy measurements [10]. Therefore, there is reason
to suggest that the spectrum may also exhibit features on small scales.

Here, we show that if the perturbation spectrum has a peak, then there is a
clear relationship between the masses of noncompact objects and PBHs; a PBH
mass of the order of $\sim10^5M_{\odot}$  corresponds to a noncompact object
mass of the
order of $\sim0.1M_{\odot}$.
For noncompact objects to be able to serve as gravitational
lenses, they must originate from sufficiently large dark matter density
fluctuations. At the radiation-dominated stage, these fluctuations
logarithmically increase with time and become nonlinear even at this stage.
To study the evolution of the fluctuations at the radiation-dominated stage,
we use the nonlinear model proposed by Kolb and Tkachev [11] and specify the
initial data for this model according to the linear theory.

Based on standard astrophysical constraints on the PBH number density in the
Universe, we obtained constraints on the rms fluctuations at the peak. In
turn, the constraints on the fluctuations give constraints on the radius of
a noncompact object. As was shown in [3], a noncompact object can serve as a
gravitational lens and can produce observable microlensing events only if
its radius exceeds the Einstein radius for this object by no more than a
factor of 10. Stringent constraints on the coefficient of nonlinear
contraction for noncompact objects follow from this condition.

The inferred relationship between noncompact objects and PBHs holds only for
adiabatic cosmological density perturbations. If the density perturbations
are entropic, then even if there are large fluctuations in the dust
component, the radiation density on the horizon scale is almost uniform and
no PBHs are formed.

It should be emphasized that when talking about a common origin of noncompact
objects and PBHs, we have in mind not the relationship between individual
noncompact objects and PBHs but the fact that the fluctuations from which
the entire collection of noncompact objects and the entire collection of
PBHs originate have a common spectrum. If a PBH emerged at some point in
space, then a noncompact object can no longer emerge at this point.
Conversely, if there is a noncompact object, then no PBH was previously (at
the radiation-dominated stage) formed at this point.

\section{THE FORMATION OF PBHs}

The PBH formation criterion was analytically derived by Carr [5] and
confirmed by numerical calculations [12, 13]. The region of space with a
density
$\rho>\rho_{\mathrm{c}}=3H^2/8\pi G$
can be roughly considered to be part of the closed Universe. Gravitational
collapse of this region and the formation of a PBH take place if the
relative radiation density fluctuation
$\delta_{\mathrm{H}}=(\rho-\rho_{\mathrm{c}})/\rho_{\mathrm{c}}$
at the time it goes under the horizon satisfies the conditions
\begin{equation}
\delta_{\mathrm{c}}\le\delta_{\mathrm{H}}\le1,\label{usl}
\end{equation}
where $\delta_{\mathrm{c}}=1/3$. The left-hand inequality implies that the
radius of the
perturbed region at the time $t$ its expansion stops exceeds the Jeans radius
$ct/\sqrt{3}$, while the right-hand inequality corresponds to the formation
of a PBH
rather than an isolated universe. The mass of the forming PBH in this model is
\begin{equation}
M_{\mathrm{BH}}=
\frac{M_{\mathrm{H}}}{3^{3/2}},\label{massa1}
\end{equation}
where $M_{\mathrm{H}}$ is the mass within the horizon.

In recent years, numerical experiments have revealed the so-called critical
gravitational collapse, during which the mass of the forming PBH is [14, 15]
\begin{equation}
M_{\mathrm{BH}}=AM_{\mathrm{H}}(\delta_{\mathrm{H}}-\delta_{\mathrm{c}})^
{\gamma},\label{critcol}
\end{equation}
where $A\sim 3$, $\gamma\simeq0.36$, and
$\delta_{\mathrm{c}}\simeq(0.65\div0.7)$. The mass
(\ref{critcol}) can be much
smaller than $M_{\mathrm{H}}$. However, as shown in [16], the PBH mass
distribution for
critical gravitational collapse is concentrated near
$M_{\mathrm{BH}}\sim M_{\mathrm{H}}$,
and the
contribution of low masses to the cosmological PBH density is modest. Here,
we consider various cases where $\delta_{\mathrm{c}}$ lies within the
range $1/3\leq\delta_{\mathrm{c}}\leq0.7$.

Following [6--8], we assume that there is a sharp peak in the fluctuation
spectrum on some fixed (in comoving coordinates) scale $\xi=r/a(t)$. Since the
PBH formation threshold is large, $\delta_{\mathrm{H}}>1/3$,
most $\xi$-scale fluctuations do
not collapse into a PBH but are preserved until the passage to the dust
stage if the dark matter particle free streaming length is moderately
large [3].
The mass $M_{\mathrm{H}}$ within the horizon as a function of the mass $M_x$ of
the dust component in fluctuations of the same comoving scale can be
calculated by using the standard Friedmann equations. Noncompact objects are
formed on time scales $t\leq t_{\mathrm{eq}}$, where
$t_{\mathrm{eq}}$ is the time of equality between
the matter and radiation densities. At this epoch, a flat cosmological model
serves as a good approximation:
\begin{equation}
\begin{array}{l}
\displaystyle{
a(\eta)=a_{\mathrm{eq}}\left[
2\frac{\eta}{\eta_*}+\left(\frac{\eta}{\eta_*}\right)^2
\right]};
\\
\\
\displaystyle{
ct=a_{\mathrm{eq}}\eta_*\left[
\left(\frac{\eta}{\eta_*}\right)^2+
\frac{1}{3}\left(\frac{\eta}{\eta_*}\right)^3
\right]},
\end{array}
\label{frid}
\end{equation}
where
$\eta_*^{-2}=2\pi G\rho_{\mathrm{eq}}a_{\mathrm{eq}}^2/3c^2$,
$a_{\mathrm{eq}}$
is the scale factor at time $t_{\mathrm{eq}}$,
\begin{equation}
\rho_{\mathrm{eq}}=\rho_0(1+z_{\mathrm{eq}})^3=3.2\cdot10^{-20}
\left(\frac{h}{0.6}\right)^{8}
\left(\frac{\Omega_{\mathrm{m}}}{0.3}\right)^{4}
\mbox{~g~sm}^{-3},
\end{equation}
$1+z_{\mathrm{eq}}=2.32\cdot10^4\Omega_{\mathrm{m}}h^2$,
$\rho_0=1.9\cdot10^{-29}\Omega_{\mathrm{m}}h^2\mbox{~g~sm}^{-3}$
is the present cosmological matter density, and $h$ is the Hubble constant (in
units of 100~km~s~$^{-1}$~Mpc~$^{-1}$). We perform our calculations for two
cosmological models: the model with the present matter density parameter
$\Omega_{\mathrm{m}}=0.3$ and the cosmological term
$\Omega_{\Lambda}=1-\Omega_{\mathrm{m}}\simeq0.7$ and the model
without the $\Lambda$ term and with $\Omega_{\mathrm{m}}=1$.
The presence of the $\Lambda$
term reduces only to a change in $\rho_{\mathrm{eq}}$;
it does not affect the formation
of noncompact objects, because the $\Lambda$ term contributes negligibly to
the total cosmological density on time scales $t\leq t_{\mathrm{eq}}$.

For $M_{\mathrm{H}}$ and $M_x$, we have the expressions
\begin{equation}
M_{\mathrm{H}}=\frac{4\pi}{3}\rho_{\mathrm{H}}(a(\eta_{\mathrm{H}})\xi)^3,~~~
M_x=\frac{4\pi}{3}\rho_0(a(\eta_0)\xi)^3.
\end{equation}
On the horizon scale, $a(\eta_{\mathrm{H}})\xi=2ct_{\mathrm{H}}$ with
$\eta_{\mathrm{H}}\ll\eta_*$ and $\rho_{\mathrm{H}}=3/32\pi
Gt_{\mathrm{H}}^2$. The present density is
$\rho_0=\rho_{\mathrm{eq}}(a_{\mathrm{eq}}/a_0)^3$.

Using (\ref{frid}), we obtain
\begin{equation}
\begin{array}{l}
\displaystyle{
M_{\mathrm{H}}=\frac{1}{2^{2/3}}
\left(\frac{3}{2\pi}\right)^{1/6}
\frac{M_x^{2/3}c}{G^{1/2}\rho_{\mathrm{eq}}^{1/6}}=}\\
\\
\qquad
\displaystyle{
=1.96\cdot10^5
\left(\frac{M_x}{0.1M_{\odot}}\right)^{2/3}
\left(\frac{\Omega_{\mathrm{m}}}{0.3}\right)^{-2/3}
\left(\frac{h}{0.6}\right)^{-4/3}M_{\odot}},
\label{mhmx}
\end{array}
\end{equation}
\begin{equation}
t_{\mathrm{H}}=\frac{GM_{\mathrm{H}}}{c^3}=
0.5
\left(\frac{M_{\mathrm{H}}}{10^5M_{\odot}}\right)\mbox{~s}.
\label{th}
\end{equation}

Denote the rms density fluctuation $\delta_{\mathrm{H}}$ by
$\Delta_{\mathrm{H}}\equiv\langle\delta_{\mathrm{H}}^2\rangle^{1/2}$.
The fraction of the radiation mass that transformed into PBHs at time
$t_{\mathrm{H}}$ is
then [5, 8]
\begin{equation}
\begin{array}{l}
{\displaystyle
\beta=
\int\limits_{\delta_{\mathrm{c}}}^{1}
\frac{\displaystyle d\delta_{\mathrm{H}}}{\displaystyle\sqrt{2\pi}\Delta_
{\mathrm{H}}}
\exp(-\frac{\displaystyle\delta_{\mathrm{H}}^2}{\displaystyle2\Delta_{\mathrm
{H}}^2})=}\\
\\
\qquad
{\displaystyle
=\frac{1}{2}\left[
{\rm erf}\left(\frac{1}{\sqrt{2}\Delta_{\mathrm{H}}}\right)-
{\rm erf}\left(\frac{\delta_{\mathrm{c}}}{\sqrt{2}\Delta_{\mathrm{H}}}\right)
\right]
\simeq
\frac{\Delta_{\mathrm{H}}}{\delta_{\mathrm{c}}\sqrt{2\pi}}\exp(-\frac{\delta_
{\mathrm{c}}^2}{2\Delta_{\mathrm{H}}^2})},
\label{bet2}
\end{array}
\end{equation}
where ${\rm erf}(x)$ is the error integral. Since, according to [16], the
fraction
of the collapsing mass of the Universe for critical gravitational collapse
is $0.8\beta$, Eq. (\ref{bet2}) for critical collapse is also a good
 approximation.

Using (\ref{mhmx}) and (\ref{bet2}), we can calculate the cosmological PBH
 density parameter
$\Omega_{\mathrm{BH}}$ at the present time $t_0$:
\begin{equation}
\begin{array}{l}
{\displaystyle
\Omega_{\mathrm{BH}}=
\frac{\beta}{2^{2/3}}
\left(\frac{3}{2\pi}\right)^{1/6}
\frac{c}{M_x^{1/3}G^{1/2}\rho_{\mathrm{eq}}^{1/6}}\simeq}\\
\\
\\
{\displaystyle
\simeq3.45\cdot10^{5}
\left(\frac{M_x}{0.1M_{\odot}}\right)^{-1/3}
\left(\frac{\Omega_{\mathrm{m}}}{0.3}\right)^{-2/3}
\left(\frac{h}{0.6}\right)^{-4/3}
\frac{\Delta_{\mathrm{H}}}{\delta_{\mathrm{c}}}\exp\left\{-\frac{\delta_
{\mathrm{c}}^2}{2\Delta_{\mathrm{H}}^2}\right\}}.
\label{nh}
\end{array}
\end{equation}

To within a factor of order unity, expression (\ref{mhmx}) can be derived from a
simple estimate
$M_{\mathrm{H}}\simeq M_xa(t_{\mathrm{eq}})/a(t_{\mathrm{H}})\simeq
M_x(t_{\mathrm{eq}}/t_{\mathrm{H}})^{1/2}$,
where
$t_{\mathrm{H}}=GM_{\mathrm{H}}/c^3$ and
$t_{\mathrm{eq}}\sim6\cdot10^{10}$~s
is the completion time of the
radiation-dominated stage. In the same way, we can derive (\ref{nh}) from the
estimate
$\Omega_{\mathrm{BH}}\simeq\beta a(t_{\mathrm{eq}})/a(t_{\mathrm{H}})$.

PBHs are formed in the tail of the Gaussian fluctuation distribution, while
most noncompact objects are formed from rms fluctuations. Therefore, we
repeat that not each fluctuation, by any means, from which a noncompact
object formed could collapse into a PBH at time $t_{\mathrm{H}}$.

\section{THE EVOLUTION OF PERTURBATIONS}

The evolution of a radiation density perturbation at the radiation-dominated
stage follows the law [17]
\begin{equation}
\delta_{\mathrm{r}}=xf(x)+\frac{3x^2}{x^2+6}\,\frac{d}{dx}f(x),\label{drdfj}
\end{equation}
where for the growing mode
$f(x)=A_{\mathrm{in}}j_1(x/\sqrt{3})$, $j_1$
is the spherical Bessel function, $A_{\mathrm{in}}$ is the normalization
constant, $x=k\eta$, and $k$ is the comoving perturbation wave vector. The
physical
perturbation wavelength satisfies the relations
\begin{equation}
\lambda_{\mathrm{ph}}(\eta)=a(\eta)\left(\frac{2\pi}{k}\right); \qquad
M_x=\frac{4\pi}{3}\rho_0
\left(\frac{\lambda_{\mathrm{ph}}(t_0)}{2}\right)^3.
\end{equation}
On the horizon scale,
$\lambda_{\mathrm{ph}}/2\simeq2ct$, $x_{\mathrm{H}}\simeq\pi$,
and we obtain from (11)
$\delta_{\mathrm{r}}=\delta_{\mathrm{H}}=A_{\mathrm{in}}\phi$,
where $\phi\simeq0.817$.

For adiabatic perturbations, the perturbation in nonrelativistic matter at
$x\ll 1$
is $\delta=3\delta_{\mathrm{r}}/4$. In [17], an analytic solution was found
for $\delta$ in the
entire interval from $x\ll1$ to  $x\gg1$ at the radiation-dominated stage.
At $x\gg1$, this solution is
\begin{equation}
\delta=\frac{3A_{\mathrm{in}}}{2}\left[
\ln\left(\frac{x}{\sqrt{3}}\right)+\gamma_{\mathrm{E}}-\frac{1}{2}
\right],\label{dgame}
\end{equation}
where $\gamma_{\mathrm{E}}-1/2\simeq0.077$ and
$A_{\mathrm{in}}=\delta_{\mathrm{H}}/\phi$ is the same as in (\ref{drdfj}).

The applicability of (\ref{dgame}) is limited to a linear stage, $\delta\ll1$.
When passing
to a nonlinear stage, we will use the spherical model from [11]. In this
model, the evolution of adiabatic perturbations is described by the equation
\begin{equation}
y(y+1)\frac{d^2b}{dy^2}+
\left[1+\frac{3}{2}y\right]
\frac{db}{dy}+\frac{1}{2}
\left[
\frac{1}{b^2}-b
\right]=0,\label{bigeq}
\end{equation}
where $y=a(\eta)/a_{\mathrm{eq}}$
and the following parametrization was introduced for the
radius of the perturbed region:
\begin{equation}
r=a(\eta)b(\eta)\xi,
\label{abxi}
\end{equation}
Here, $\xi$ is the comoving coordinate and
$b(\eta)$ allows for the deceleration
of cosmological expansion in the region of enhanced density. The quantity
$b$
in (\ref{abxi}) can be expressed in terms of $\delta$ as
\begin{equation}
b=(1+\delta)^{-1/3}.\label{delb}
\end{equation}
This relation means the passage from the Eulerian description of the
perturbation evolution (\ref{dgame}) to its Lagrangian description (\ref
{bigeq}).

To solve (\ref{bigeq}) requires specifying $\delta_{\mathrm{i}}$ at some
 initial
$y_{\mathrm{i}}$,
$b_{\mathrm{i}}=(1+\delta_{\mathrm{i}})^{-1/3}$ according to (16),
and the expansion rate $db/dy$. In [11], Eq.(\ref{bigeq})
was solved for entropic perturbations when the initial velocity may be
disregarded, $db/dy\simeq0$. In our case of adiabatic perturbations,
the initial
velocity is large; we specify it according to the solution (\ref{dgame}). At
$x\gg1$
and $y\ll1$, we have
\begin{equation}
\begin{array}{l}
\displaystyle{
x=\frac{\pi}{2^{2/3}}\left(\frac{3}{2\pi}\right)^{1/6}
\frac{yc}{M_x^{1/3}G^{1/2}\rho_{\mathrm{eq}}^{1/6}}=}\\
\qquad
\\
\displaystyle{
=2.86\cdot10^6y
\left(\frac{\Omega_{\mathrm{m}}}{0.3}\right)^{-2/3}
\left(\frac{h}{0.6}\right)^{-4/3}
\left(\frac{M_x}{M_{\odot}}\right)^{-1/3}}.\label{xy}
\end{array}
\end{equation}
We obtain from (\ref{dgame}), (16), and (17)
\begin{equation}
\left.\frac{db}{dy}\right|_{y_{\mathrm{i}}}=-\frac{\delta_{\mathrm{H}}b_
{\mathrm{i}}^4}
{2y_{\mathrm{i}}\phi}.
\end{equation}
We solve Eq. (\ref{bigeq}) numerically. The time
$y_{\mathrm{i}}$ must be chosen in such a way
that (\ref{dgame}) and (\ref{bigeq}) are simultaneously valid.
An optimum choice is the time
when $\delta_{\mathrm{i}}=0.2$. A test shows that the
results of our calculations change by
no more than 15\%
for a different choice of $\delta_{\mathrm{i}}$ in the range
$0.1-0.3$. Having
specified $\delta_{\mathrm{i}}$, we obtain
$x_{\mathrm{i}}$ and
$y_{\mathrm{i}}$ from (\ref{dgame}) and (\ref{xy}). The evolution of
$\delta=b^{-3}-1$ is illustrated in Fig.~1.

The cosmological expansion of noncompact objects stops when
$dr/dt=0$.
According to [11], this is equivalent to
\begin{equation}
\frac{db}{dy}=-\frac{b}{y}\label{ost}.
\end{equation}
Denote $b$ and $y$ at the time the expansion stops by
$b_{\mathrm{max}}$ and $y_{\mathrm{max}}$,
respectively. The CDM density in noncompact objects at the same time is
\begin{equation}
\rho_{\mathrm{max}}=\rho_{\mathrm{eq}}y_{\mathrm{max}}^{-3}
b_{\mathrm{max}}^{-3}
\end{equation}
and, consequently, the radius of the noncompact object at the stoppage time is
\begin{equation}
R_{\mathrm{max}}=\left(\frac{3M_x}{4\pi\rho_{\mathrm{max}}}
\right)^{1/3}.\label{rmax}
\end{equation}
The numerically calculated dependence of $R_{\mathrm{max}}$ on
$\delta_{\mathrm{H}}$ is shown in Fig.~2.

Thus, we have shown how the radius of a noncompact object at the time its
cosmological expansion stops can be determined for the specified
perturbation $\delta_{\mathrm{H}}$ on the horizon scale.

\section{THE PARAMETERS OF \\ GRAVITATIONAL LENSES}

After its cosmological expansion stops, a noncompact object begins to
contract, with its final radius being
\begin{equation}
R_x=\kappa R_{\mathrm{max}},
\end{equation}
where
$\kappa$ is the coefficient of nonlinear contraction. It is generally
believed [18] that after the cosmological expansion of a noncompact object
stops, it is virialized, radially contracting by a factor of 2, i.e.,
$\kappa=0.5$.
According to the theory of gravitational instability,
$\kappa\simeq0.3$ in a
multiflow region [3]. At present, the physical processes that could cause a
noncompact object to contract to
$\kappa<0.3$ are unknown. Therefore, we take
$\kappa\simeq0.3$ as the lower limit.

For a noncompact object to be able to serve as a gravitational lens producing
observable microlensing events in the Galactic halo, its radius should not
significantly exceed the Einstein radius
\begin{equation}
R_{\mathrm{E}}=2\sqrt{GM_xd/c^2},\label{rensh}
\end{equation}
where
$d\sim20$~kpc for microlensing in the halo. We define
$\varepsilon$ as
\begin{equation}
\varepsilon\equiv10\frac{R_{\mathrm{E}}}{R_x}\label{vardef}
\end{equation}
According to calculations [3], the inequality
$\varepsilon\ge1$ must hold. If $\varepsilon<1$, then
the theory comes into conflict with observational data on the light curves
[3]. One of the microlensing events with a lens mass
$M_x\simeq0.02M_{\odot}$ was studied
in detail in [19]. A comparison of the observed and calculated light curves
showed that if a noncompact object has no baryonic core at its center, then
it must have the size
$R_x=1.6\cdot10^{13}$~cm and a compactness parameter (in
our notation)
$\varepsilon\simeq19$. In the presence of a baryonic core with a
mass of
$0.05M_x$,
$R_x=5.7\cdot10^{13}$~cm and
$\varepsilon\simeq4.8$. It was noted in
[19] that the model of a point-like lens for this event is also acceptable
because of the large observational errors.

We obtain from (\ref{rmax}), (\ref{rensh}), and (\ref{vardef})
\begin{equation}
\kappa\varepsilon=10\frac{R_{\mathrm{E}}}{R_{\mathrm{max}}}=
\frac{1.9\cdot10^{15}\mbox{~cm}}{R_{\mathrm{max}}}
\left(\frac{M_x}{M_{\odot}}\right)^{1/2}
\left(\frac{d}{20\mbox{~kpc}}\right)^{1/2}
\end{equation}
On the other hand, using our calculations (see the preceding section), we
derived a relationship between
$R_{\mathrm{max}}$ and
$\delta_{\mathrm{H}}$; the quantity
$\delta_{\mathrm{H}}$ defines the
present cosmological PBH density according to Eq. (\ref{nh}). It should be noted
that noncompact objects are formed from rms fluctuations. Therefore,
$\delta_{\mathrm{H}}$ from
the preceding section should be set equal to the rms fluctuation
$\delta_{\mathrm{H}}=\Delta_{\mathrm{H}}$.
Using our numerical calculations, we derive the dependence of
$\Omega_{\mathrm{BH}}$ on the
product $\kappa\varepsilon$ (see Figs.~3 and 4).

There are several astrophysical constraints on the mass and number of PBHs.
It follows from a constraint on the age of the Universe that
$\Omega_{\mathrm{BH}}\le1$. If
PBHs provide the dominant part of dark matter in the Galaxy, then they must
tidally interact with globular clusters by disrupting them. The PBH mass was
constrained for this case in [20],
$M_{\mathrm{BH}}\le10^4M_{\odot}$. At
$\Omega_{\mathrm{BH}}\sim1$, PBHs are
capable of distorting the CMB spectrum if they are formed about 1~s after
the annihilation of
$e^+e^-$- pairs [5]. Mass accretion by black holes at the
pre-galactic and present epochs contributes to the background radiation in
different wavelength ranges. However, calculations strongly depend on the
model and yield
$\Omega_{\mathrm{BH}}\le10^{-3}\div10^{-1}$ for
$M_{\mathrm{BH}}\sim10^5M_{\odot}$. In [22], the
constraint
$\Omega_{\mathrm{BH}}<0.1$ on intergalactic PBHs was
obtained from the condition
for the absence of reliable gamma-ray-burst lensing events for
$10^5M_{\odot}<M_{\mathrm{BH}}<10^9M_{\odot}$.
A more stringent lensing constraint,
$\Omega_{\mathrm{BH}}<0.01$ for the mass range
$10^6M_{\odot}<M_{\mathrm{BH}}<10^8M_{\odot}$,
was obtained from VLBI observations of compact radio
sources [23].

Let us first consider the microlensing event studied in [19] by assuming
that $\varepsilon=0.3$. If the noncompact object has no
baryonic core (the vertical
line
$\kappa\varepsilon\simeq5.7$ in Figs.~3 and 4 corresponds to this case), then
the constraint
$\Omega_{\mathrm{BH}}<0.1$ rules out the interpretation of this event as
microlensing by noncompact objects for all
$\delta_{\mathrm{c}}=1/3\div0.7$ in the two
cosmological models under consideration. In the presence of a baryonic core
(the vertical line
$\kappa\varepsilon\simeq1.44$), the case with
$\delta_{\mathrm{c}}=0.7$ remains admissible in the
cosmological model without the
$\Lambda$ term (Fig.~4).

Consider the less stringent condition $\kappa\varepsilon>0.3$, which is
satisfied at
$\varepsilon>1$
and
$\kappa>0.3$. At
$\Omega_{\Lambda}\simeq0.7$ (Fig.~3), the constraint
$\Omega_{\mathrm{BH}}<0.1$ rules out all
models with
$\delta_{\mathrm{c}}<0.5$. At
$\Omega_{\mathrm{BH}}>10^{-6}$, the case with
$\delta_{\mathrm{c}}\simeq0.7$ remain possible.
If $\Omega_{\Lambda}=0$, then the constraint
$\Omega_{\mathrm{BH}}<0.1$ leaves a narrow region (see Fig.~4)
with
$M_x>0.1M_{\odot}$ and
$\delta_{\mathrm{c}}=1/3$. The constraint
$\Omega_{\mathrm{BH}}<10^{-3}$ completely rules out
the models with
$\delta_{\mathrm{c}}=1/3$, but allows the cases with
$\delta_{\mathrm{c}}>1/3$.

\section{CONCLUSIONS}

We have shown that the formation of noncompact dark-matter objects
(neutralino stars) proposed in [2, 3] to account for the observed
microlensing events in the Galactic halo must be preceded by the formation
of a PBH with a mass of the order of
$\sim10^5M_{\odot}$. These PBHs and neutralino
stars are formed from the same peak in the primordial fluctuation spectrum.

Astrophysical constraints on the number of PBHs in the Universe allowed us to
constrain the coefficient of nonlinear contraction and compactness parameter
for a noncompact object
$\kappa\varepsilon$; our constraints are shown in Figs.~3 and 4. The
most stringent constraints are obtained in the presence of a cosmological
term
$\Omega_{\Lambda}\simeq0.7$. In this case,
noncompact objects can serve as gravitational
lenses only at a large PBH formation threshold,
$\delta_{\mathrm{c}}>0.5$, which was
calculated in the model of critical gravitational collapse. At smaller
$\delta_{\mathrm{c}}$,
the model of noncompact objects as microlensing objects is ruled out. The
constraint on
$\kappa\varepsilon$ is significantly relaxed if the
$\Lambda$ term is small
(Fig.~4). However, this possibility is currently considered to be unlikely.

It is important to note that to avoid the situation where
$\ln(\Omega_{\mathrm{BH}})\ll0$
requires an accurate adjustment of the parameters for the inflationary model
that ensures that
$\Delta_{\mathrm{H}}$ is in a narrow interval,
$\simeq0.05\div0.12$. Therefore, if the
observed microlensing events are actually produced by noncompact objects,
then these objects most likely originate from entropic density
perturbations. In this case, our constraints are removed, as discussed in
the Introduction.

The constraints can also be significantly relaxed if for some reason, PBHs
are formed in smaller quantities than that given by expression (\ref{bet2}).
 This is
possible, for example, in the case of great importance of the nonlinear
effects that accompany the generation of metric perturbations at the
inflationary stage if these effects suppress the appearance of large
fluctuations [24]. Note, however, that the role of nonlinear effects is
presently not completely understood, and the results of calculations depend
on the specific inflationary model. For example, it was found in [25] that
nonlinear effects cause the PBH formation probability to increase, which is
directly opposite to the result from [24].

Conversely, if the noncompact nature of lenses will be proven in the future,
then for adiabatic density perturbations, this leads us to conclude that a
large number of PBHs with masses of the order of
$\sim10^5M_{\odot}$ can be formed at
the radiation-dominated sage of the Universe. These PBHs must affect the
evolution of galaxies and their nuclei. It may well be that one of such PBHs
was found by the Chandra space X-ray observatory in the galaxy M 82 [26]. In
a separate paper [27], we develop a model for the formation of galaxies
through multiple mergers of protogalaxies with condensation centers in the
form of massive PBHs. The mergers of galaxies and the growth of central
massive black holes at the galactic nuclei take place simultaneously with
the formation of galaxies. The recently found correlations between the
masses of central black holes and the bulge velocity dispersion have been
shown to naturally arise in this scenario.

Since noncompact objects and PBHs originate from a common perturbation
spectrum, we can in principle reconstruct the shape of the perturbation
spectrum and determine the PBH mass function from the distribution of
noncompact objects in mass and radius using a Press--Schechter-type theory
[28]. Unfortunately, only a few objects were detected by microlensing, and
such a calculation will become possible only in the future when the
statistics will improve.

We may consider a situation that, in a sense, is reverse to the situation
described previously. According to the hypothesis [7], the dark halo objects
responsible for microlensing are PBHs with masses of the order of
$\sim(0.01\div1)M_{\odot}$.
If elementary particles with a free streaming length
$\ll \xi$ comprise the
remaining part of the dark matter, then miniclusters will be formed from
these particles at the dust stage. Using formula (\ref{mhmx}), we can
 immediately
estimate the minicluster mass as
$\sim10^{-11}\div10^{-8}M_{\odot}$. Such masses are
possible if the mass of the dark-matter particles exceeds 1 GeV [3]. The
formation of PBHs with
$\Omega_{\mathrm{BH}}\sim1$ requires rms fluctuations
$\Delta_{\mathrm{H}}\simeq0.06$. At time
$t_{\mathrm{eq}}$, fluctuations in the dust component
$\delta\sim1$, which give rise to
miniclusters, correspond to them.

\bigskip

ACKNOWLEDGMENTS

\medskip

We thank K.P. Zybin and V.N. Lukash for helpful discussions. This study was
supported by the INTAS (grant no. 99-1065) and the Russian Foundation for
Basic Research (project nos. 01-02-17829, 00-15-96697, and 00-15-96632).

\newpage

FIGURE CAPTIONS

\bigskip

Fig. 1. The evolution of a CDM density perturbation
$\delta$. The plot
corresponds to the parameters
$\delta_{\mathrm{H}}=0.04$, $M_x=0.1M_{\odot}$, and
$\Omega_{\mathrm{m}}=0.3$. The
curve was obtained from formula (\ref{dgame}) before the point
($y_{\mathrm{i}}=3.96\cdot10^{-6}$, $\delta_{\mathrm{i}}=0.2$),
and a numerical solution of Eq. (\ref{bigeq}) was used at
$y>y_{\mathrm{i}}$. The
cosmological expansion of a noncompact object stops at the
radiation-dominated stage at
$y=a/a_{\mathrm{eq}}\simeq0.49$. The dashed line indicates
the evolution of
$\delta$ according to the linear theory (\ref{dgame}).

\medskip

Fig. 2. Radius
$R_{\mathrm{max}}$ of noncompact objects at the time their expansion stops
versus perturbation magnitude
$\delta_{\mathrm{H}}$ on the horizon scale. Solid curves 1, 2,
and 3 correspond to the masses of noncompact objects
$M_x=1M_{\odot}$, $0.1M_{\odot}$, and
$0.01M_{\odot}$ in the cosmological model with
$\Omega_{\mathrm{m}}=0.3$. Dashed lines 4, 5, and 6
were constructed for
$\Omega_{\mathrm{m}}=1$ at the same masses.

\medskip

Fig. 3. PBH density parameter
$\Omega_{\mathrm{BH}}$ versus nonlinear contraction coefficient
and compactness parameter for noncompact objects
$\kappa\varepsilon$ in the model with
$\Omega_{\mathrm{m}}=0.3$.
Curves (1, 2, 3) correspond to the masses of noncompact objects
$M_x=(0.01, 0.1, 1)M_{\odot}$ at
$\delta_{\mathrm{c}}=1/3$.
Curves (4, 5, 6) correspond to
$\delta_{\mathrm{c}}=1/2$ and
curves (7, 8, 9) correspond to
$\delta_{\mathrm{c}}=0.7$
for the same masses. The horizontal and vertical lines correspond to the
upper observational limits
$\Omega_{\mathrm{BH}}=0.1$,
$\Omega_{\mathrm{BH}}=10^{-3}$ and to
$\kappa\varepsilon=0.3$, $1.44$,
respectively. The admissible regions lie below the horizontal lines and to
the right from the vertical lines.

\medskip

Fig. 4. Same as Fig. 3 for the cosmological model with
$\Omega_{\mathrm{m}}=1$. The vertical lines correspond to
$\kappa\varepsilon=0.3$, $1.44$ and $5.7$.

\newpage

\begin{figure}
\includegraphics{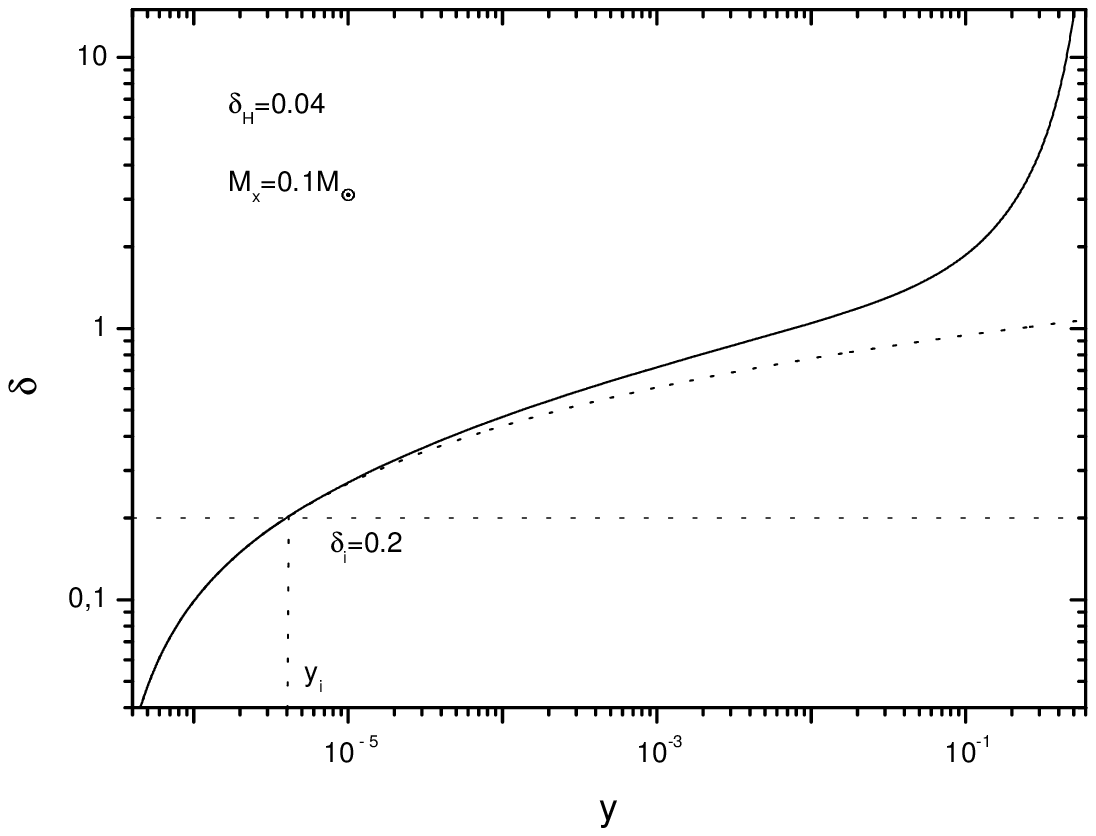}
\label{fig1}
\end{figure}

\newpage

\begin{figure}
\includegraphics{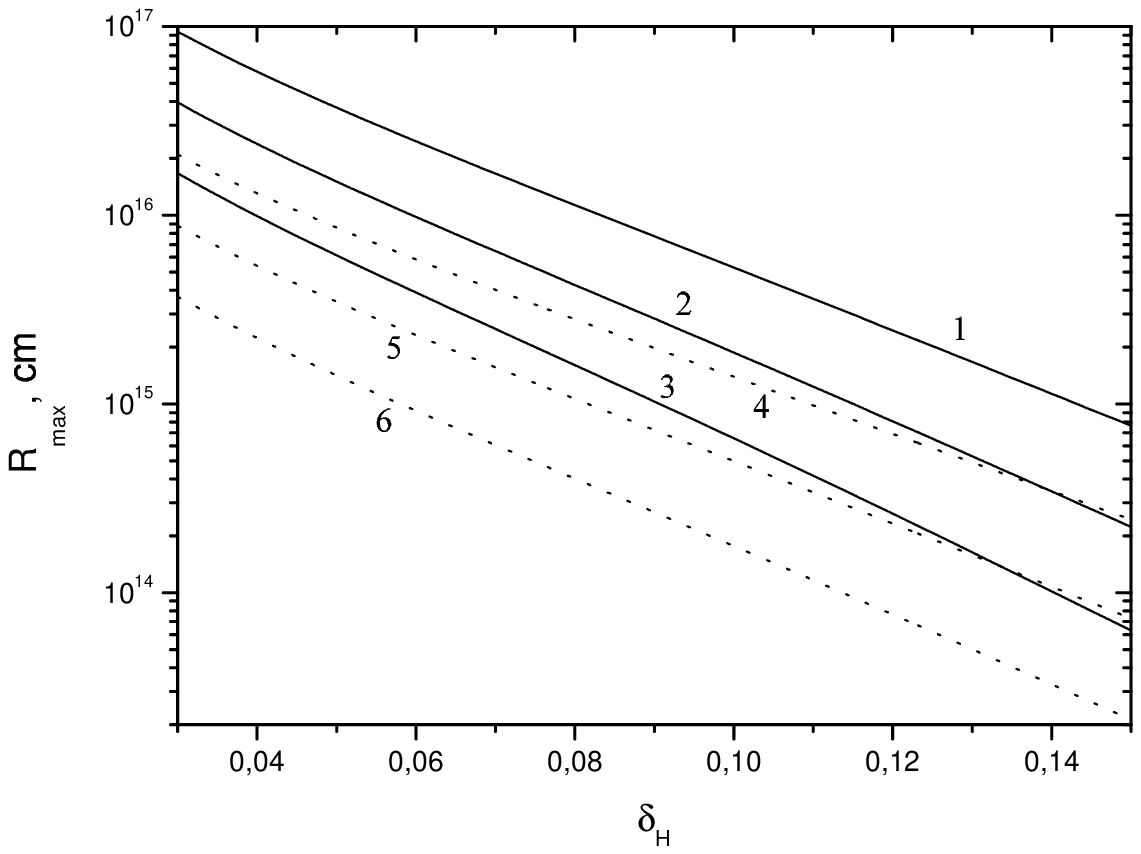}
\label{fig2}
\end{figure}

\newpage

\begin{figure}
\includegraphics{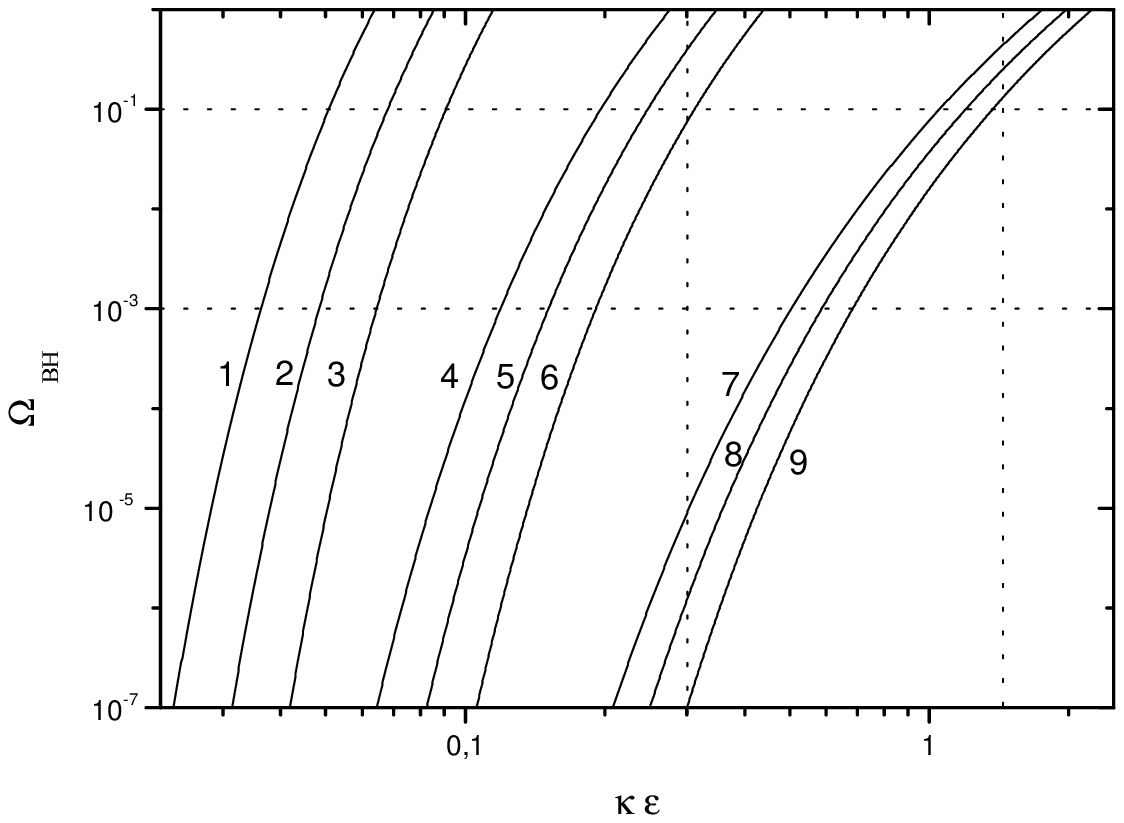}
\label{fig3}
\end{figure}

\newpage

\begin{figure}
\includegraphics{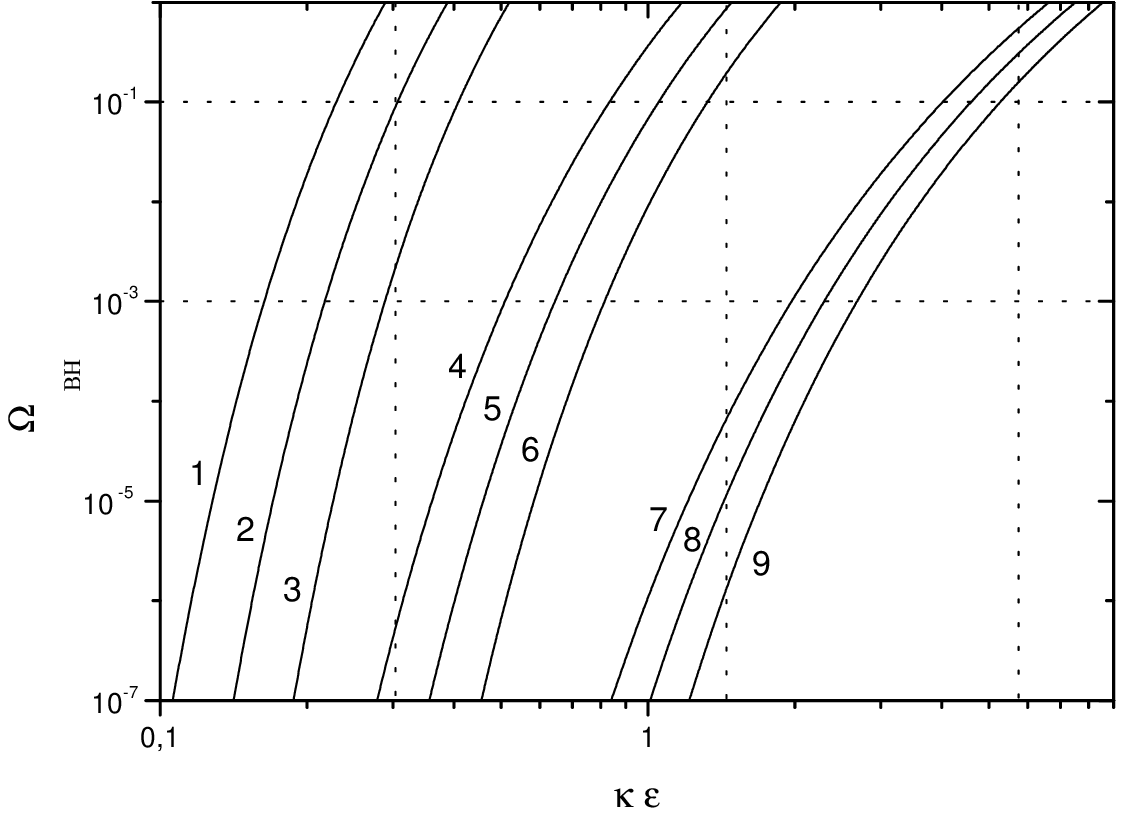}
\label{fig4}
\end{figure}


\begin{thebibliography}{99}

\bibitem{alc}
C. Alcock, R. A. Allsman, D. R. Alves, T. S. Axelrod, A. C. Becker, D. P.
Bennett, K. H. Cook, N. Dalal, A. J. Drake, K. C. Freeman, M. Geha, K. Griest,
M. J. Lehner, S. L. Marshall, D. Minniti, C. A. Nelson, B. A. Peterson,
P. Popowski, M. R. Pratt, P. J. Quinn, C. W. Stubbs, W. Sutherland, A.
B. Tomaney, T. Vandehei, and D. Welch,  Astrophys. J. {\bf 542},  281
(2000).
\bibitem{gz951}
A. V. Gurevich and K. P. Zybin,
Phys. Lett. A {\bf208},  276 (1995).
\bibitem{ufn2}
A. V. Gurevich, K. P. Zybin and V. A. Sirota, Physics-Uspekhi {\bf40}, 869
(1997), E-print archive  astro-ph/9801314.
\bibitem{zeld67}
Ya. B. Zel'dovich and I. D. Novikov,
Sov. Astron. {\bf10},  602 (1967).
\bibitem{carr75}
B. J. Carr,
Astrophys. J. {\bf201}, 1 (1975).
\bibitem{star}
A. A. Starobinsky, JETP Lett.
{\bf55}, 489 (1992).
\bibitem{ivan94}
P. Ivanov, P. Naselsky, and I. Novicov,
Phys. Rev. D {\bf50}, 7173 (1994).
\bibitem{yok95}
J. Yokoyama, E-print archive astro-ph/9509027.
\bibitem{gars96}
J. Garcia--Bellido, A. Linde, and D. Wands,
Phys. Rev. D {\bf54}, 6040 (1996).
\bibitem{apm}J. Barriga, E. Gaztanaga, M. G. Santos,
S. Sarkar, E-print archive  astro-ph/0011398.
\bibitem{kt}
E. W. Kolb, and I. I. Tkachev, Phys. Rev. D {\bf50}, 769 (1994).
\bibitem{nad78}
D. K. Nadezhin, I. D. Novikov and A. G. Polnarev
Sov. Astron.
{\bf22}, 129 (1978).
\bibitem{nov79}
I. D. Novikov, A. G. Polnarev, A. A. Starobinsky, and Ya. B. Zeldovich,
Astron. Astrophys. {\bf80}, 104 (1979).
\bibitem{chop} M. W. Choptuik, Phys. Rev. Lett. {\bf70}, 9, (1993).
\bibitem{niem98}
J. C. Niemeyer, E-print archive  astro-ph/9806043.
\bibitem{yok98}
J. Yokoyama, E-print archive  astro-ph/9804041.
\bibitem{ssw}
C. Schmid, D. J. Schwarz, and P. Widerin, Phys. Rev. D {\bf59}, 043517
(1999).
\bibitem{sasl}
W. C. Saslaw, {\it Gravitational Physics of Stellar and Galactic
Systems} (Cambridge Univ. Press, Cambridge, 1985).
\bibitem{sir} V. A. Sirota, JETP, {\bf90}, 227 (2000).
\bibitem{moor}
B. Moore, Astrophys. J. Lett. {\bf413},  93 (1993).
\bibitem{carr79}
B. J. Carr, MNRAS {\bf189}, 123 (1979).
\bibitem{nemir}
R. J. Nemiroff, G. F. Marani, J. P. Norris, and J. T. Bonnel,
Phys. Rev. Lett {\bf86}, 580 (2001).
\bibitem{wilk}
P. N. Wilkinson, D. R. Henstock, I. W. A. Browne, A. G. Polatidis, P. Augusto,
A. C. S. Readhead, T. J. Pearson, W. Xu, G. B. Taylor, and R. C. Vermeulen,
Phys.  Rev. Lett. {\bf86}, 584 (2001).
\bibitem{nelin} J. S. Bullock, and J. B. Primack,
E-print archive  astro-ph/9806301.
\bibitem{ivan97}P. Ivanov, E-print archive  astro-ph/9708224.
\bibitem{kaar}
P. Kaaret, A. H. Prestwich, A. Zezas,
S. S. Murray, D. W. Kim,  R. E. Kilgard, E. M. Schlegel, and
M. J. Ward, E-print archive  astro-ph/0009211.
\bibitem{we1}V. I. Dokuchaev and Yu. N. Eroshenko,
Astron. Lett.
{\bf27}, 759 (2001).
\bibitem{bugaev}
E. V. Bugaev, and K. V. Konishchev, E-print archive  astro--ph/0103265.

\end{thebibliography}
\end{document}